\documentclass[final,5p,times,twocolumn]{elsarticle}

\usepackage{graphicx}
\usepackage[english]{babel}
\usepackage[utf8]{inputenc}
\usepackage[T1]{fontenc}

\usepackage{amssymb}
\usepackage{amsfonts}
\usepackage{amsmath}
\usepackage{color}
\newcommand{\citer}[1]{Ref.\cite{#1}}
\newcommand{\citers}[1]{Refs.\cite{#1}}

\journal{}

\begin{document}

\begin{frontmatter}



\title{Contribution to the angular momentum transport paradigm for accretion disks}


\author[a1,a2]{Giovanni Montani}
\author[a1,a3]{Nakia Carlevaro}
\address[a1]{ENEA, Fusion and Nuclear Safety Department, C.R. Frascati,\\ Via E. Fermi 45, 00044 Frascati (Roma), Italy}
\address[a2]{Physics Department, ``Sapienza'' University of Rome,\\ P.le Aldo Moro 5, 00185 Roma, Italy}
\address[a3]{CREATE Consortium,\\ Via Claudio 21 (80125) Napoli, Italy}

\begin{abstract}
We analyze the stationary configuration of a thin axisymmetric stellar accretion disk, neglecting non-linear terms in the plasma poloidal velocity components. We set up the Grad-Shafranov equation for the system, including the plasma differential rotation (according to the so-called co-rotation theorem). Then, we study the small scale backreaction of the disk to the central body magnetic field and we calculate the resulting radial infalling velocity. We show that the small scale radial oscillation of the perturbed magnetic surface is associated to the emergence of relevant toroidal current densities, able to balance the Ohm law even in the presence of quasi-ideal values of the plasma resistivity. The contribution to the infalling velocity of the averaged backreaction contrasts accretion, but it remains negligible as far as the induced magnetic field is small compared to that of the central body.
\end{abstract}



\end{frontmatter}



\section{Introduction}
One of the most intriguing open questions in astrophysics concerns the angular momentum transport across accreting structures. 
The basic idea, formulated in \citer{Sh73} for stellar thin accretion disks, relies on the crucial role played by shear viscosity 
associated to the disk differential rotation 
(for a review on this topic, see \citers{BH98,BKL01,R07}). However, it was clear since the very beginning that such viscosity effects could not have origin form the binary interaction. In fact, in the 
typical range of density and temperature available for thin stellar disks, the addressed ion-electron plasma is quasi-ideal.
In order to account for the plasma viscosity requested by the observed accretion rates, it was originally postulated (and still argued in the literature) that turbulence is responsible for dissipative effects. 

The origin of the plasma turbulent behavior was identified in the Magneto-Rotational Instability (MRI), see \citer{Ve59,Ch60,BH91}, which is an Alfv\'enic mode emerging in the presence of a 
weak magnetic tension (see also \citer{Ba03}). Although the presence of this instability in stellar accretion disks appears rather natural, the presence of a magnetic field implies that the Ohm law must be included in the plasma description. Moreover, it is not still settled down the idea that MRI can also generate significant values of the plasma resistivity. Since the currents present in the disk are only those ones
provided by the plasma backreaction to the central body magnetic field (very small if the induced magnetic 
field has the same spatial scale of the equilibrium), the balance of 
the Ohm law requires large value of the disk resistivity, especially 
in the X-ray binary systems where the accretion rate is particularly relevant.
The necessity of an \emph{anomalous resistivity} is a rather puzzling question of the so-called Standard Model for thin accretion disks \cite{Sh73}.

In \citer{Co05}, it was shown that, if the $\beta$-parameter (i.e. the ratio of the plasma pressure to the magnetic pressure) of the disk 
plasma is large enough, then the backreaction has a much smaller scale 
with respect to the equilibrium one and a ring-like current emerges on 
the radial profile. In the limit of an induced magnetic field larger than the pre-existing one (that one of the central body), the 
disk can also be decomposed into a ring-like structure, since the matter density acquires periodic radial nodes \cite{CR06}. 
This scenario has been extended to a global radial profile in \citer{MB11pre}, where the microscopic nature of the radial profile oscillations is 
also clearly stated. 

In \citer{MC12}, it was firstly argued that the emergence of a microscopic magnetic structure in the disk could be associated with the balance 
of the Ohm law without the request of an anomalous resistivity.
In fact, the small scale backreaction can be associated with relevant 
values of the plasma current densities. However, in the framework of a reduced one-dimensional model, it has been outlined an inconsistency between the standard accretion theory and the so-called crystalline morphology of the disk \cite{Co05}. This inconsistency is here revised and overcome in a more general two-dimensional scheme and by introducing the concept of the spatial average on the microscales.
 
In this paper, we consider a two-dimensional axisymmetric thin disk configuration in which the differential rotation is included in the Grad-Shafranov 
equation (GSE), while the poloidal velocities, in view of their smallness, are 
retained linearly only in the system describing the global plasma equilibrium. We fix the detailed features of the plasma disk backreaction to the central body magnetic field, demonstrating that the inconsistency mentioned above can be overcome and that the crystalline morphology of the disk allows to balance the Ohm law. In fact, the 
radially oscillating current density generates a 
correspondingly oscillating radial component of the velocity 
(in principle greater than the Standard Model contribution), but it 
averages to zero on a macroscopic scale, i.e. on the background profile scale.

\section{Comparison to similar previous approach}

It is worth to briefly discussing the nature of the crystalline structure of the disk in comparison to the Standard Model of accretion and also to better clarify the novelty of the present analysis with respect to similar previous studies, especially that one provided in \citer{MC12}.

The new issue addressed by the original analysis in \cite{Co05}, which first introduced the crystalline morphology of the disk, consists in observing that the steady configuration of the plasma can include a non-negligible backreaction, since for sufficiently large value of the $beta$-parameter, it acquires a small scale structure. By other words, while the deviation of the magnetic flux function from its background morphology is small, such a correction has relevant gradient and therefore it can be responsible for significant current densities in the disk. This fact marks a significant difference between this formulation and the Standard Model, in which there is a single relevant spatial scale and the current densities due to the backreaction of the plasma disk to the magnetic field of the central object are expected to be small. 

The resulting picture of the crystalline profile is characterized by a large scale magnetic field of the background configuration, here essentially coinciding with the dipole field of the central object, plus a magnetic contribution, which radially oscillates on a smaller spatial scale. In \cite{CR06}, this model has been extended to the case of a non-linear response of the plasma, i.e. there the backreaction magnetic field exceeds the background one and then it is possible to observe a small scale fragmentation of the plasma disk into a ring-like profile. While in \cite{Co05,CR06} the construction refers to a local model, valid near enough to a fiducial radius, in \citer{MB11pre} the global profile of the radial oscillation of the magnetic configuration is properly described.

In \citer{MC12}, it was moreover outlined that, if the disk plasma has sufficiently high $\beta$-parameter value, then an inconsistency seems to emerge between the crystalline morphology and the Shakura formula for the mass accretion rate. The considered model was a reduced one-dimensional one, based on an averaging procedure on the disk vertical depth. In this approximation, the dependence of the
differential rotation on the magnetic flux function is, to some extent an ad hoc assumption, justified by the Ferraro theorem \cite{Fe37}, which holds for a stationary (two-dimensional) axisymmetric configuration, as discussed in some detail in Appendix A.

The reason of the emerging inconsistency was identified in the discrepancy between the Shakura mass accretion rate, provided by the azimuthal momentum conservation equation and that one evaluated by the Ohm law. In fact, the latter has a source due to the backreaction current density, having the form of oscillating terms (linear in the second gradients of the perturbed magnetic flux function).

In the present study, we propose as a solution to this inconsistency, the observation that, when averaged on a large radial scale (like that one of the background) this current provides a vanishing contribution. On the same level, the average of quadratic terms, due to the contribution of the backreaction to the mass accretion rate, can provide, on average, a finite non-vanishing contribution. It is just this additional term, neglected in the study developed in \citer{MC12}, which can provide a zero average of the left-hand side of the Ohm law.

This compensation effect could be discussed even in a reduced one-dimensional model, since the vertical gradients are always smaller than the radial ones, but we consider a real two-dimensional axisymmetric disk, in order to pursue a more general analysis and, in particular, to deal with a Grad-Shafranov-like equation for the disk equilibrium, with the natural assumption that the radial and vertical velocity fields be small enough to neglect their quadratic contributions (in \citer{MC12} it 
was constructed an hydro-dynamical equilibrium on a different scale with respect to that one of the crystalline radial profile).

\section{Basic Formalism}
We consider an axisymmetric thin accretion 
disk ($H/r\ll1$, being $r$ the radial distance 
from the center of symmetry and 
$H(r)$ the slowly varying half-depth), 
embedded in the gravitational and magnetic 
field of the central object, 
having mass $M$ and magnetic dipole moment $\mu$. Adopting standard cylindrical coordinates 
$(r,\,\phi,\,z)$, the central object gravitational field 
is described by the potential 
$\phi_G = - G_N M/(r^2 + z^2)^{1/2}$ ($G_N$ being the 
Newton constant), while the magnetic field is 
fixed by the dipole flux function 
$\psi_D = \mu r^2/(r^2 + z^2)^{3/2}$, which is a valuable approximation at sufficient large distances from the central body.

In this Section, we discuss the equilibrium configuration of the thin disk 
as described by the visco-resistive magneto-hydrodynamic (MHD) theory, under the 
following (rather natural) assumptions \cite{Sh73}: 
\begin{itemize}
\item[i)] the viscosity and resistivity coefficients are taken 
constant over the studied disk region, this assumption does not affect the general character of our analysis but simplifies the analytical treatment and it can be always regarded as a good approximation in a sufficiently small radial interval of the disk;
\item[ii)] the poloidal components of the plasma velocity are much smaller than the azimuthal one (associated to the disk differential 
rotation) and therefore they can be neglected in the radial 
and vertical equilibria, since they would appear quadratically;
\item[iii)] we address a purely poloidal magnetic field in agreement to the dipole field of the central object, neglecting dynamo effects.
\end{itemize}

Thus, we consider a velocity field of the plasma disk 
having the form
\begin{equation} 
\textbf{v} = \textbf{v}_p + \omega r\hat{\textbf{e}}_{\phi}
\, , 
\label{acd1}
\end{equation}
where $\textbf{v}_p = v_r\hat{\textbf{e}}_r + v_z \hat{\textbf{e}}_z$ and $\omega$ denotes the non-uniform angular velocity of the disk. 
Once assigned the magnetic flux function $\psi$, the 
magnetic field inside the disk reads
\begin{equation} 
\textbf{B} = -\frac{1}{r}\partial _z\psi \hat{\textbf{e}}_r + 
\frac{1}{r}\partial _r\psi \hat{\textbf{e}}_z 
\,. 
\label{acd2}
\end{equation}
Since this field is associated to a purely azimuthal 
current density, the corresponding component 
of the Ohm law provides the equation 
\begin{equation} 
\textbf{v}_p\cdot {\nabla}\psi = \frac{c}{4\pi \sigma}D\psi 
\, , 
\label{acd31}
\end{equation}
where $c$ is the speed of light, $\sigma = const.$ denotes the electric conductivity coefficient, $\textbf{v}_p\cdot {\nabla} \equiv v_r\partial _r 
+ v_z\partial _z$ 
and  $D\psi = \partial _r^2\psi + 
\partial _z^2\psi -(1/r)\partial _r\psi$. 
The poloidal component of the same equation  
provides an electric field of the form 
\begin{equation}
\textbf{E} = - \omega{\nabla}\psi/c\,,
\end{equation}
whose irrotational character (we deal with a steady configuration) 
implies $\omega = \omega (\psi)$, according to the 
so-called co-rotation theorem \cite{Fe37} (see also Appendix A).

Neglecting the quadratic terms in the poloidal velocities, 
the radial and vertical equilibria are associated to the following  
system of equations:
\begin{equation} 
\omega ^2 r\hat{\textbf{e}}_r = \frac{1}{\rho}{\nabla}p 
+ {\nabla}\phi_G + \frac{1}{4\pi \rho r^2}D\psi 
{\nabla}\psi 
\,, 
\label{acd3} 
\end{equation}
where $\rho$ and $p$ are the plasma mass density and 
pressure, respectively, 
and we neglected 
the disk self-gravity. 

We now assume that the disk is isothermal on each 
magnetic surface, i.e. we take the disk temperature as 
$T = T(\psi)$. Hence, the first and second thermodynamical 
principles easily provide the basic relation
\begin{equation} 
\frac{1}{\rho}{\nabla}p = {\nabla}G + S\frac{dT}{d\psi} 
{\nabla}\psi 
\, , 
\label{acd4} 
\end{equation} 
where $G$ and $S$ are the Gibbs function and the entropy 
per unit mass, respectively. 
Substituting this expression into the system (\ref{acd3}), 
we arrive to a Grad-Shafranov-like equation for the 
disk equilibrium, see \citer{Ogilvie97}, i.e. 
\begin{equation} 
D\psi = - 4\pi \rho r^2 \left( \frac{d\mathcal{E}}{d\psi} + S\frac{dT}{d\psi} 
+ r^2\omega \frac{d\omega}{d\psi}\right) 
\, , 
\label{acd5}
\end{equation} 
where 
\begin{equation} 
\mathcal{E} \equiv G + \phi_G - \omega ^2r^2 /2
\, 
\label{acd6_} 
\end{equation}
is the Bernoulli-like function and we used the 
proportionality ${\nabla}\mathcal{E}\propto{\nabla}\psi$ 
to infer the relation $\mathcal{E} = \mathcal{E} (\psi)$. 

It can be checked that the 
azimuthal component of the equilibrium, 
taking into account the co-rotation 
theorem, can be expressed in the form 
\begin{eqnarray}
\textbf{v}_p\cdot {\nabla}\psi 
+ \frac{2\omega}{rd\omega /d\psi} v_r = 
\frac{\eta}{\rho}  
	\left( D\psi + \frac{4}{r}\partial_r\psi\right) + 
	\nonumber \\
	+ \frac{\eta}{\rho d\omega /d\psi}\frac{d^2\omega}{d\psi^2}\left[
\left( \partial _r\psi\right)^2 + \left(\partial _z\psi\right)^2\right] \,,
\label{acd6} 
\end{eqnarray}
where $\eta = const.$ is the 
shear viscosity coefficient. 

Finally, the mass conservation equation, 
which reads
\begin{equation}
\frac{1}{r}\partial _r\left( \rho rv_r\right) + \partial _z\left( \rho v_z\right) = 0
\, , 
\label{acd8x}
\end{equation}
admits the solution 
\begin{equation}
\textbf{v}_p = -\frac{1}{\rho r}
\partial_z\Theta \hat{\textbf{e}}_r + 
\frac{1}{\rho r}\partial _r\Theta 
\hat{\textbf{e}}_z
\, , 
\label{acd9x}
\end{equation}
$\Theta (r,z)$ denoting a generic function.

\subsection{Basic idea of the proposed scheme}\label{subsubsub}
One of the crucial equations concerning the present analysis and its comparison to the standard model for accretion is the azimuthal component of the Ohm law, i.e. Eq.(\ref{acd31}) that, for convenience we rewrite here explicitly as:
\begin{equation}
v_zB_r - v_rB_z = J_{\phi}/\sigma
\, . 
\label{naprd1}
\end{equation}
If we indicate by the suffix $0$ and $1$ the background and the perturbed quantities, respectively, and by $L$ and $\lambda$ the corresponding spatial scales, in the Shakura model we have:
\begin{equation}
v_{r0}\frac{\psi_{0}}{L}\sim \frac{c}{4\pi \sigma}\frac{\psi_{1}}{\lambda^2}
\, . 
\label{naprd2}
\end{equation}
The value of $v_{r0}$ is provided by the azimuthal component of the momentum conservation and it is fixed by the plasma viscosity. Since $|\psi_{1}|\ll|\psi_{0}|$ and, in the Shakura model $\lambda \sim L$, 
to get consistency between the momentum and the Ohm azimuthal components, the conductivity coefficient $\sigma$ must be properly modeled and, in 
some situations (like in X-ray binary systems), it must acquire an \emph{anomalous} small value \cite{MP13}.

Here, we analyze a very different situation: in the correspondence to a large value of the plasma $\beta$-parameter, the backreaction lives on a scale $\lambda \ll L$ and the function $\psi_{1}$ radially oscillates on such a small scale \cite{Co05}. 
In this case, the radial infalling velocity has, in addition to the standard contribution $v_{r0}$, a dominant contribution $v_{r1}\propto \psi_{1}/\lambda^2$, radially oscillating on the scale $\lambda$. 
Since $\psi_{1}\sim \sin[2\pi r/\lambda]$, $v_{r1}$ clearly vanishes when spatially averaged on the background scale $L$. 
This consideration solves phenomenologically the inconsistence rised in \citer{MC12}, where it was outlined the dominant character of the oscillating radial velocity. 

The crucial point, here, is that also 
the current density $J_{\phi}\equiv J_{\phi1}$ is proportional to 
$\psi_{1}/\lambda^2$, so that it averages to zero on the macroscopic scale $L$. Thus, also the left-hand-side of Eq.(\ref{naprd1}) must vanish and the value of $\sigma$ is no longer involved in the problem. 
The vanishing average on the scale $L$ of such a left-hand-side is due to the compensation of the contribution 
$v_{r0}B_{z0}$ with the quadratic terms in the backreaction, which do not average to zero (they behave like $\sim \sin^2[2\pi r/\lambda]$ and average to $1/2$). 
We conclude by noting that the balance of these two contributions requires 
the following relation of magnitude orders:
\begin{equation}
\psi_{0}^2/L^3\sim\psi_{1}^2/\lambda^3\, . 
	\label{naprd3}
\end{equation}
The validity of this condition must be thought as a constraint on the allowed values of the backreaction amplitude $\psi_{1}$, since $\lambda$ is determined by the background features, as we will discuss in the next Section.

We observe that, in the present model, there is no contribution of the background to the right-hand side of Eq.(\ref{acd31}), because our background magnetic field exactly coincides with that one of the central body (in particular, we will deal with 
a dipole profile). Therefore, in the disk region, it is a vacuum field, i.e. we must have $D\psi_0 = 0$. However, it is worth noting that, more in general, Eq.(\ref{naprd3}) implies the following relation
\begin{equation}
J_0 \sim \frac{\psi_0}{L^2} \sim 
\frac{\psi_1}{\lambda^2}
\frac{\psi_1L}{\psi_0\lambda} \sim 
J_1 \frac{B_1}{B_0}
\, , 
\label{14xx}
\end{equation}
where $J_0$ and $J_1$ denote the background and perturbed current density magnitudes, respectively. Thus we see that, as far as we remain in the linear limit, i.e. $B_1\ll B_0$, the contribution of the background to the current density in the right-hand side of Eq.(\ref{acd31}) remains always negligible (even if not exactly zero like here).

\section{Linear Plasma Backreaction}
The background configuration (we denot $\psi_0=\psi_D$) is
fixed via the relations
\begin{equation}
{\nabla}\mathcal{E}_0 + r^2\omega _0
{\nabla}\omega _0 = 0
\,, \quad \; \frac{dT_0}{d\psi _D} = 0
\, ,  
\label{xacd10}
\end{equation}
which correspond to the following
isothermal gravostatic equilibrium
\begin{equation}
{\nabla}p_0 = \rho_0 \Big(
\frac{1}{2}\omega _0^2 {\nabla}r^2 -
{\nabla}\phi_G\Big)
\, ,
\label{xacd11}
\end{equation}
where we recall that the label $0$ refers to background quantities. The addressed isothermal assumption ($T(\psi)\equiv T_0=const.$), provides $p_0 = v_s^2 \rho_0$,
where the sound speed $v_s$ is constant. The
vertical component of Eq.(\ref{xacd11}) gives the
following density profile
\begin{equation}
\rho _0\simeq \bar{\rho}(r)\Big(1 - \frac{z^2}{2H^2}\Big)
\, ,
\label{acd12}
\end{equation}
where $\bar{\rho}$ denotes the mass density profile on the equatorial plane, $H\equiv v_s/\omega_K$ ($\omega_K \equiv \sqrt{G_NM/r^3}$
being the Keplerian angular velocity) and we assumed to deal
with a thin disk, i.e. $H/r\ll 1 \Rightarrow r\ll L_s\equiv G_NM/2v_s^2$. Finally, 
in order to fix the form of the function $\omega (\psi)$,
we use the relation 
between $\omega_K$ and $\psi_D$, 
valid on the equatorial plane,
extrapolating it in the whole thin disk configuration,
namely
\begin{equation}
\omega (\psi ) = \psi ^{3/2}\sqrt{G_NM/\mu^3}
\, .
\label{acd13}
\end{equation}

We now show that, in the absence of the plasma 
disk backreaction, the system of Eqs.(\ref{acd31}) and (\ref{acd6}) is 
actually inconsistent. 
In fact, since the dipole magnetic field 
is a vacuum solution of the Maxwell equations, 
we have $D\psi_D = 0$. 
Then, from Eq.(\ref{acd31}), we 
get the condition 
$\textbf{v}_p\cdot {\nabla} \psi_D = 0$, 
which, by means of Eq.(\ref{acd9x}), 
gives the simple relation $\Theta = 
\Theta (\psi_D )$. 
As a consequence, since $\partial_z\Theta \propto 
\partial_z\psi_D$, we have $v_r=0$ on the equatorial plane $z=0$. The inconsistency is with Eq.(\ref{acd6}) which provides, via simple algebra and using Eq.(\ref{acd13}), $v_r=v_r^{Sh}$, where 
\begin{align}
v_r^{Sh}\equiv - 21\eta /8\rho r
\end{align}
is just the standard Shakura value, apart from the assumption $\eta = const.$

In the standard model for accretion, this inconsistency is 
solved by including in Eq.(\ref{acd31}) the current density due to 
the plasma backreaction $\propto D \psi_1$, where the magnetic 
flux function is now given by $\psi = \psi_D + \psi_1$ (with $|\psi_1|\ll|\psi_D|$). However, if $\psi_1$ and $\psi_D$ have a comparable spatial scale of variation, the compatibility of 
the accretion model requires a very low value of the 
electric conductivity $\sigma$, not always justified by the kinetic 
conditions of the accreting plasma. 
To avoid this difficulty, we pursue a different point of 
view, based on requiring that the spatial
scale of variation of $\psi_1$ be much smaller than that of the 
background (as discussed in Sec. \ref{subsubsub}).

Assuming that in Eq.(\ref{acd5}) the plasma backreaction weakly affects the functions 
$\mathcal{E}(\psi)$ and using Eq.(\ref{acd13}), we can arrive to 
a relevant representation for the linear response of the disk to the 
presence of the central object magnetic field. In fact, defining 
\begin{eqnarray}
	\psi = \psi_D + \psi_1 \equiv \psi _D + \varphi\sqrt{r} \, , 
	\label{acd14}\\ 
	k_0^2 \equiv 4\pi \frac{3G_N M\bar{\rho} r^3}{\mu^2} = 
	3\frac{\omega _K^2}{v_A^2}
	\, , 
	\label{acd15}
\end{eqnarray} 
where $v_A$ denotes the Alfv\'en velocity on the equatorial plane, for a thin disk and requiring $k_0r\gg 1$ 
(to deal with a small scale backreaction), 
we get the following perturbed equation for $\varphi$
\begin{equation}
\partial ^2_r\varphi + \partial ^2_z\varphi
= - k_0^2\Big( 1 - \frac{z^2}{2H^2}\Big)\varphi
\, . 
\label{acd16}
\end{equation}
This equation admits the solution 
\begin{equation}
\varphi = A\sin (kr)\, \exp \Big[
-\frac{k_0z^2}{2\sqrt{2}H}\Big]
\,, 
\label{acd16bis}
\end{equation}
where, for the sake of simplicity, we omitted a phase term (clearly a combination of sine and cosine is still a solution) and it can be shown how $k\equiv k_0 (1-1/\sqrt{3\beta})$ (here $\beta=2v_s^2/v_A^2$) and $A$ is an integration constant (with $A\sqrt{r}\ll |\psi _D|$).

This solution has been also discussed in \citer{MB11pre}, 
generalizing the analyses in \citers{Co05,CR06} to a global 
radial profile. The difference is that there the obtained 
perturbed magnetic flux function corresponds to a weak backreaction of the plasma, while here we could also deal with the case in which 
the induced magnetic field is comparable to the central 
object one (in the limit in which $\mathcal{E}(\psi)$ is not significantly perturbed). In fact, here we only implemented the smallness of 
$\psi_1$ and neglected the mass density fluctuations with respect to 
the background value $\rho_0$, without requiring the smallness of the backreaction magnetic field.

\section{Physical Considerations and Conclusions}

We now observe that
$\partial^2_r\psi\simeq \partial^2_r\psi_1 \sim -k^2\psi_1$,
but the macroscopic spatial average (on the background scale) of this
quantity clearly vanishes and the same behavior concerns the vertical
derivatives. By other words, on a macroscopic level, the induced
magnetic field and current density are on average
(in the following denoted by $\langle ... \rangle$) zero and only quadratic terms in the
backreaction, like those ones appearing in Eq.(\ref{acd6}), survive.
Thus, averaging Eq.(\ref{acd31}), we get
$\langle \textbf{v}_p\cdot {\nabla}\psi \rangle = 0$.
Substituting this basic result into the averaged version of Eq.(\ref{acd6}),
we arrive to the following expression for the
radial infalling velocity
\begin{equation} 
v_r=v_r^{nl}\equiv v_r^{Sh}\Big[ 1 - 
\frac{k_0^2A^2}{14\mu^2}r^5\Big(1+ \frac{1}{3\beta}\Big)\Big]
\, ,
\label{fire}
\end{equation}
where we made use of the expressions
\begin{equation}
\langle (\partial _r\psi_1 )^2\rangle = k_0^2A^2r/2\;,\quad 
\langle (\partial _z\psi_1 )^2\rangle = k_0^2A^2r/6\beta\,.
\end{equation}

Clearly, the non-linear correction appearing in the radial   
velocity (\ref{fire}) is positive (i.e. associated to matter radial 
ejection) and it is of the same order of the Shakura infalling 
contribution when the following condition holds
\begin{equation}
\frac{Ar^{3/2}}{\mu}\gtrsim \frac{1}{k_0r}
\;\;\Rightarrow \;\;
|\partial_r\psi_1|\gtrsim |\partial_r\psi_D|\, , 
\label{arr}
\end{equation}
which implies a plasma backreaction vertical magnetic 
field comparable or greater than that of the central body
(this extreme situation has been considered in \citer{CR06}, 
arguing the possible decomposition of the disk into a 
ring microstructure). 

Let us now discuss how the inconsistency presented in \citer{MC12} can be solved, i.e. how the condition $\langle \textbf{v}_p\cdot \nabla\psi\rangle = 0$ is compatible with the expression for the radial velocity in Eq.(\ref{fire}). The key point here is easily identified if we observe that the radial velocity receives, in Eq.(\ref{acd6}), the most important contribution by the term $D\psi = D\psi_1$. This term averages macroscopically to zero, but it gives an important non-vanishing contribution when it is first multiplied by $\partial_r\psi_1$ and then macroscopically averaged. It is just this additional term that has to compensate the zeroth-order term $v_r^{Sh}\partial_r\psi_D$ to provide $\langle \textbf{v}_p\cdot \nabla\psi\rangle=0$. However, to get this cancellation, the term coming from the backreaction must be of the same order of the background one (see the condition (\ref{naprd3})), more specifically we must have
\begin{equation}
\psi_1/\psi_D \sim 1/(kr)^3 \, . \label{accc} 
\end{equation}
It is important to stress how the condition above lives in the linear regime of backreaction, as it immediately comes out by the comparison with the inequality in Eq.(\ref{arr}) (we recall that $kr\gg1$) which characterizes the emergence of a non-linear regime. By other words, the consistency (averaging on microscales) of Eq.(\ref{acd31}) and Eq.(\ref{acd6}) fixes the ratio between $\psi_1$ and $\psi_D$, once the scale of the backreaction is assigned via background quantities.

We now address the following point of view standing 
as the fundamental conclusion of our analysis. 
If, as it is rather natural in many real situations, the thin disk plasma has a sufficiently large value of the $\beta$-parameter, 
the backreaction is of small scale and, even when the induced 
magnetic field is small (i.e. $v^{nl} \simeq v^{Sh}$), 
the emerging poloidal current is relevant in magnitude and 
it plays a crucial role in the Ohm law. 
In fact, now the quantity $\textbf{v}_p\cdot {\nabla}\psi$ 
is no longer zero, like for the background, but it vanishes on a macroscopic average.
The existence of such a rapidly oscillating current along 
the disk radial profile can ensure the balance of the Ohm law 
even in the presence of a kinetic (non-anomalous) plasma 
resistivity. In the limit of a strong non-linear backreaction 
of the plasma, as discussed above, the accretion feature of the disk can be 
suppressed up to favoring radial material ejection.

Furthermore, we observe that the crystalline profile of the disk is generated by the balance of the Lorentz force in the disk, i.e. the $j\times B$ term with the perturbation of the centripetal force as effect of the co-rotation theorem, when the magnetic flux function is perturbed (we recall that a current density exists only as a consequence of the disk backreaction to the central body magnetic field). Since the magnetic flux function is radially oscillating as a result of the equilibrium between these two terms, the same behavior characterizes the current density. In practice, we can say that the disk radial distribution is endowed with concentrically shaped rings of current, two adjacent ones being oriented in opposite directions. This current rings are locally relevant, but when a large scale average is considered, they cancel out each another and the net macroscopic current density is essentially zero.

It is worth noting that, on the equatorial plane where the magnetic field
is essentially vertical, the condition for dealing with MRI can be written as \cite{BH98,MCP16,CM17}
$k_{MRI}^2v_A^2 < 3\omega _K^2$. Using the definition of $k_0$ given in Eq.(\ref{acd15}), this expression simply states $k_{MRI} < k_0$. In the proposed scheme, the MRI is an efficient instability at the macroscopic
scales of the disk, responsible for the turbulent viscosity, while the
microscopic backreaction is able to ensure the balance of the Ohm law.
By other words, the original Shakura idea is not
affected by the magnetic microstructures, which play an important role
in determining the real current density toroidally living in the disk. Clearly, there is a window of $\beta$ values, for which the MRI is still triggered, but
the microscale backreaction is suppressed and, in this
situation, an anomalous value of the resistivity
is mandatory for dealing with the observed angular
momentum transport in rapidly accreting sources,
like X-ray binary stars. We also observe that the possibility to have significant accretion features, in correspondence to small resistivity value of the plasma, has the consequence to reduce the diffusive character of the magnetic field via the disk, which an anomalous resistivity would unavoidably imply in the MHD induction equation. Actually, it has been demonstrated that a strong diffusion of the magnetic field prevents that it takes enough large values in the disk to trigger the jet formation (see for instance \citer{2000ApJ529978B,BisnovatyiKogan:2007p4253,Sp08}).

\section*{Appendix A}  
We demonstrate here the validity of the co-rotation theorem under the assumptions at the ground of the analysis of the present manuscript. We start by observing that, for a steady configuration, the electric field receives no contribution from the vector potential and therefore it reduces to the gradient of a scalar potential, i.e. we must ensure its irrotational character: 
\begin{equation}
\nabla\times \textbf{E} = 0 = \nabla\times \left(-\textbf{v}\times \textbf{B} + c\textbf{J}/\sigma \right) /c\, ,  
\label{a1}
\end{equation}
where, we made use of the Ohm law.

In the correspondence to the poloidal magnetic field
\begin{equation}
\textbf{B} = -  \frac{1}{r}\partial_z\psi \hat{\textbf{e}}_r 
+ \frac{1}{r}\partial_r\psi\hat{\textbf{e}}_z
\label{a2}
\end{equation}
and to the velocity field
\begin{equation}
\textbf{v} = \textbf{v}_p + \omega r\hat{\textbf{e}}_{\phi}
\, , 
\label{a3}
\end{equation}
the $\phi$-component of Eq.(\ref{a1}) can be expressed as 
\begin{equation}
\partial_r\omega\partial_z\psi - \partial_z \omega\partial_r\psi = 0
\, . 
\label{a4}
\end{equation}
It is the necessity to reduce this equation to an identity, which lead to the validity of the co-rotation theorem, namely $\omega = \omega(\psi)$. The poloidal component of Eq.(\ref{a1}) is naturally reduced to an identity by virtue of the azimuthal component of the Ohm law, namely Eq.(\ref{acd31}). 

Finally we observe that, from the Faraday equation, the relation 
$\partial_tB_{\phi} = - c(\nabla\times \textbf{E})_{\phi} = 0$ holds. 
Thus, if, as it is natural, the central object magnetic field has a negligible azimuthal component, it can not be generated by a dynamo effect.


\end{document}